%% file: main.tex
\documentclass[preprint]{vgtc}               

\ifpdf
  \pdfoutput=1\relax                   
  \usepackage{graphicx}                
  \DeclareGraphicsExtensions{.pdf,.png,.jpg,.jpeg} 
\else
  \usepackage{graphicx}                
  \DeclareGraphicsExtensions{.eps}     
\fi%

\graphicspath{{figures/}{pictures/}{images/}{./}} 

\usepackage{microtype}                 
\PassOptionsToPackage{warn}{textcomp}  
\usepackage{textcomp}                  
\usepackage{mathptmx}                  
\usepackage{times}                     
\usepackage{cite}                      
\usepackage{tabu}                      
\usepackage{booktabs}                  

\onlineid{0}

\vgtccategory{Research}

\vgtcinsertpkg

\preprinttext{IEEE VR 2019 Workshop on Human Augmentation and its Applications}

\title{wavEMS: Improving Signal Variation Freedom of Electrical Muscle Stimulation}

\author{Michinari Kono\thanks{e-mail: mchkono@acm.org}\\ %
     \scriptsize The University of Tokyo %
\and Jun Rekimoto\thanks{e-mail: rekimoto@acm.org}\\ %
     \parbox{1.4in}{\scriptsize \centering The University of Tokyo \\ Sony Computer Science Laboratories, Inc.}}

\abstract{There has been a long history in electrical muscle stimulation (EMS), which has been used for medical and interaction purposes. Human-computer interaction (HCI) researchers are now working on various applications, including virtual reality (VR), notification, and learning. For the electric signals applied to the human body, various types of waveforms have been considered and tested. In typical applications, pulses with short duration are applied, however, many perspectives are required to be considered. In addition to the duration and polarity of the pulse/waves, the wave shapes can also be an essential factor to consider. A problem of conventional EMS toolkits and systems are that they have a limitation to the variety of signals that it can produce. For example, some may be limited to monophonic pulses. Furthermore, they are usually limited to rectangular pulses and a limited range of frequencies, and other waveforms cannot be produced.  These kinds of limitations make us challenging to consider variations of EMS signals in HCI research and applications. The purpose of ``{\it wavEMS}'' is to encourage testing of a variety of waveforms for EMS, which can be manipulated through audio output. We believe that this can help improve HCI applications, and to open up new application areas.%
} 

\CCScatlist{
  \CCScatTwelve{Human-centered computing}{Human computer interaction (HCI)}{Interaction devices}{Haptic devices};
  \CCScatTwelve{General and reference}{Document types}{General literature}{};
  \CCScatTwelve{Human-centered computing}{Human computer interaction (HCI)}{Interaction devices}{}
}

\begin{document}


\firstsection{Introduction}

\maketitle

\input{body}

\acknowledgments{
We thank Marutsuelec and the developers of the amplifier for giving us suggestions for improvement. }

\bibliographystyle{abbrv-doi}

\bibliography{bib2019}
\end{document}

%% file: body.tex
The use of electrical muscle stimulation (EMS) is becoming popular in human-computer interaction (HCI) and virtual reality (VR) fields. While the original technique is rooted in the medical field for rehabilitation purposes, it is now common to explore interactive applications. Before HCI researchers found their interest for EMS, some artists used them for embodied performances. In 1995, an artist Stelarc used electrical stimulation for his performance~\cite{stelarc}. Furthermore, Elsenaar~\cite{facialhacking} and Manabe~\cite{facevisualizer} applied EMS on to their face for their performances. Following these applications, in 2010, Tamaki et al.~\cite{phand2010} introduced EMS to HCI by presenting finger manipulation techniques, which was a pioneering work in the field. Since this work, many researchers are now using EMS for HCI, and many workshops have been held (e.g.,~\cite{emsstudio}).

While EMS can perform as a strong mean for human augmentation purposes, researchers may still find it difficult to understand the appropriate waveforms and safety concerns. These kinds of knowledge tend to spread around various research fields, which makes HCI researchers difficult to follow them comprehensively~\cite{tochiguide}. In HCI, researchers mainly use square waves for the stimulation, however, using other waveforms may have potential to expand the effect of the electrical stimulation. Therefore, to further encourage researchers to explore using various waveforms, we provide literature on the effect of EMS waveforms and a system that can provide various waveforms for the stimulation. The system allows to output signals through audio input, where the audio signals are directly converted to electrical stimulation. We named this system {\it wavEMS}, which is a wireless mobile EMS toolkit.

The contributions of this paper, are concluded as follows.
\begin{itemize}
    \item We conclude the history and the effect of waveforms of EMS from literature in research fields other than HCI, and to help researchers consider exploring various waveforms of EMS.
    \item We propose {\it wavEMS}, which is a toolkit for researchers and practitioners. The system can be controlled via wireless audio signals, and the waveforms can be designed freely.
\end{itemize}

\section{Research on EMS in HCI}
\subsection{Application in HCI}
There are many EMS research for various applications, including VR~\cite{impacto,emswall}, learning~\cite{phand2010,emspercussion}, and many others. Muscle-plotter~\cite{muscleplot} used multiple channels of EMS to guide users to move their hand for hand output activities. Niijima et al.~\cite{faceemslimb} applied EMS to the face, in order to control the contraction of the bodily muscles. Pfeiffer et al.~\cite{pedems} applied EMS on to the user's leg to control the walking direction. We can notice that EMS can be applied to various body parts. This is also an important feature of EMS, where they can be built in a tiny package with strong output force~\cite{tinyems}. Another interesting domain is to use EMS for emotion expression~\cite{inpulse,emotionactuator}. As well as Impacto~\cite{impacto}, which is a work that used EMS for VR feedback, the tactile element (tingling) of EMS can disturb the performance of such applications. We believe that there is a possibility where we can restrain this negative effect by considering more suitable EMS waveforms.

\subsection{Systems for EMS in HCI}
To build applications, researchers have also developed some systems for EMS. Let your body Move toolkit~\cite{letbodymove} is a common system that is used for many following EMS research in HCI. Zap++~\cite{zap} was a system with a maximum of 20 channel output. The system is designed for a sleeve worn on the arm, where various combinations of the electrodes are allowed for the stimulation output. {\it multi-ems}~\cite{ahems} was also an opensource toolkit for multiple channel output. For these systems, the signals of the EMS are controlled by switches, and generate square waved pulses for output. Manabe et al. used an EMS board for their workshop~\cite{bodyhack}, which was controlled by audio output. The board also allows improving the signal variations of EMS. However, it consists of two large transformers (140~g each), which can disturb its mobility. Another concern is that where the system can allow dangerous signal output when used without appropriate knowledge. Therefore, we provide a user interface for popular waveforms to restrict dangerous output, and literature for effective signals. Another benefit of {\it wavEMS} is where it can be small/light and can be controlled via wireless communication.

\section{History and Parameters of Electrical Muscle Stimulation}
\begin{figure}[h]
\centering
  \includegraphics[width=1.00\columnwidth]{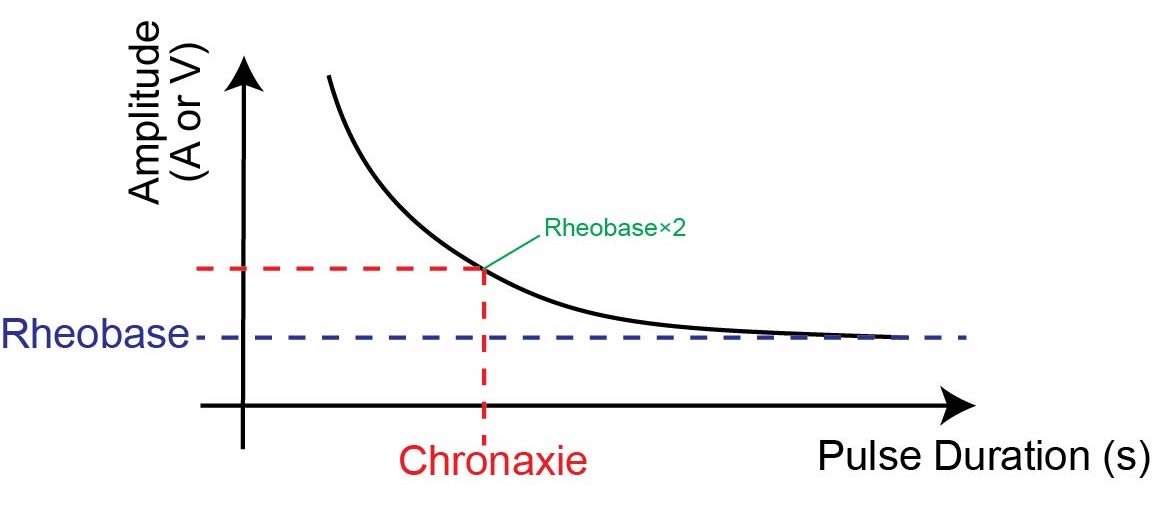}
  \caption{Chronaxie and rheobase are points on the amplitude-duration curve for stimuling tissues/nerves.}
      \label{fig:chronaxie}
\end{figure}

The history of EMS is rooted from 1791, when Luigi Galvani used electric current to cause a muscle of a frog to constract~\cite{galvanilife}. At this time the mechanism was still unknown, however, attracted great scientific interest, and was studied through the 1800s to reveal the principles~\cite{emsdigest}. In 1903, Prof. Leduc introduced a form of electric current to minimize the energy for muscle contraction~\cite{responseleduc}. This was where the pulsed inputs that are still used nowadays, were investigated and applied for muscle stimulation. For the experiments, an anaesthetized cat was used to observe the effect of stimulation~\cite{responseleduc}. The terms ``Chronaxie'' and ``Rheobase'' was first stated in Lapicque's paper in 1909~\cite{chronaxie100}. These terms are used to discuss the relationship between the strength and duration of a rectangular pulse and it's threshold (Figure~\ref{fig:chronaxie}). Later in 1977, ``Russian currents'' have been reported by Yakov Kots~\cite{russianes}. Russian currents consist of 2.5 kHz AC signals that are burst modulated at 50 Hz with a 50 \% duty cycle. Although this seemed to be an effective method for electrical stimulation, it is now been studied to be inferior to other types of signals~\cite{emsdigest}.  Since then, various research with various signals has been conducted and may devices have been created for rehabilitation and training purposes~\cite{reviewems1986, reviewemsdevice}.

\begin{figure}[h]
\centering
  \includegraphics[width=1.00\columnwidth]{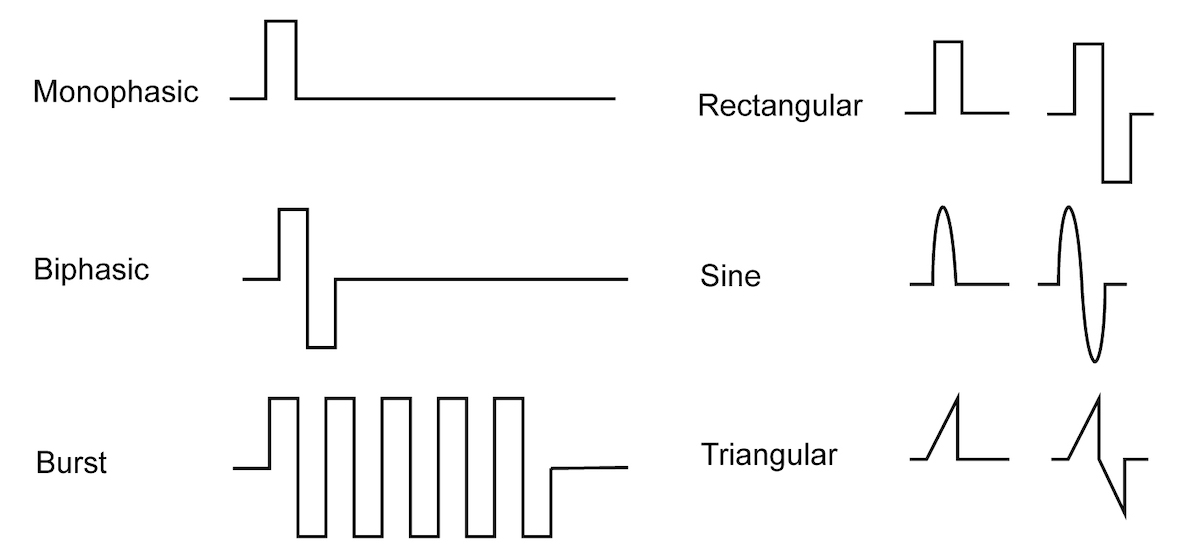}
  \caption{Types of waveforms used in EMS. A specific stimulus, burst sine wave of a carrier frequency of 2,500 Hz and burst and inter-burst duration of 10 ms is called a Russian current. }
      \label{fig:waves}
\end{figure}

When using EMS, considering the type of signal is an important element. There are studies that have compared various types of waveforms~\cite{reviewemsdevice,nemswaves} (Figure~\ref{fig:waves}). Ara\'{u}jo et al. noted that square waveforms can be more uncomfortable compared with triangular and quadratic waveforms~\cite{nemswaves}. Laufer et al. said that monophasic and biphasic signals have an advantage over polyphasic waveforms (2.5 kHz) regarding the generated fatigue~\cite{fatiguenems3}. Petrofsky et al. compared sine, square and the Russian waveform, and found that sine waves were less painful, still having a greater muscle strength~\cite{waveformems}. For monophasic and biphasic signals, there are arguments for both waveforms~\cite{facialhacking,apptensems,textileems}. For example, biphasic waves can give a sharper sensation, while they can be benefical to reduce fatigue. Prausnitz presented a review for the effect of the current applied to the skin~\cite{effectcurrentskin}. For interactive use cases, Knibbe et al. concluded that various parameters can be explained differently when experienced~\cite{expems}. Other considerations include the electrode size that also may influence the stimulation of EMS~\cite{electrodesize}. Researchers should carefully consider the parameters for their purposes.

\section{wavEMS}

\subsection{Hardware Design}
\begin{figure}[h]
\centering
\includegraphics[width=1.0\columnwidth]{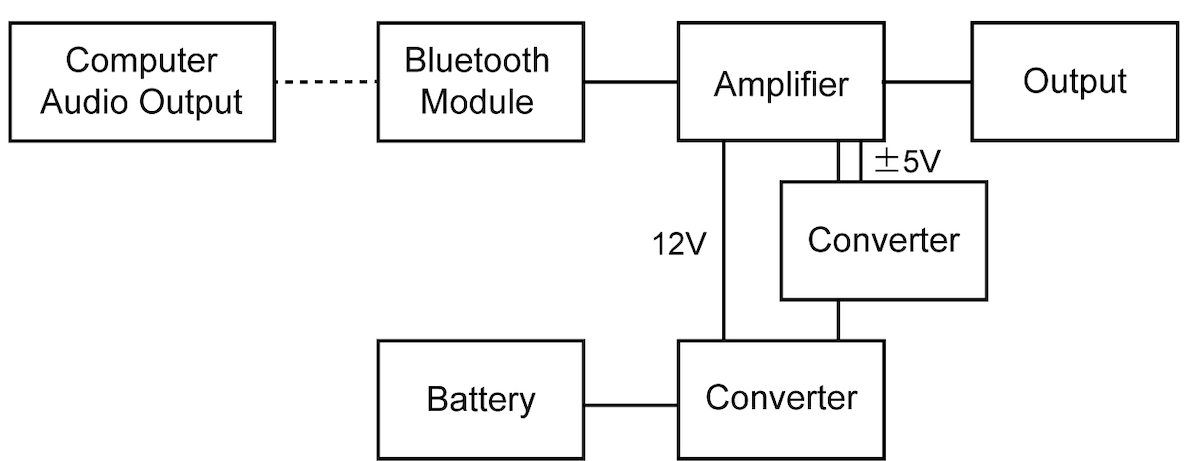}
\caption{The system overview of {\it wavEMS}. }
\label{fig:system}
\end{figure}

\begin{figure}[h]
\centering
\includegraphics[width=1.0\columnwidth]{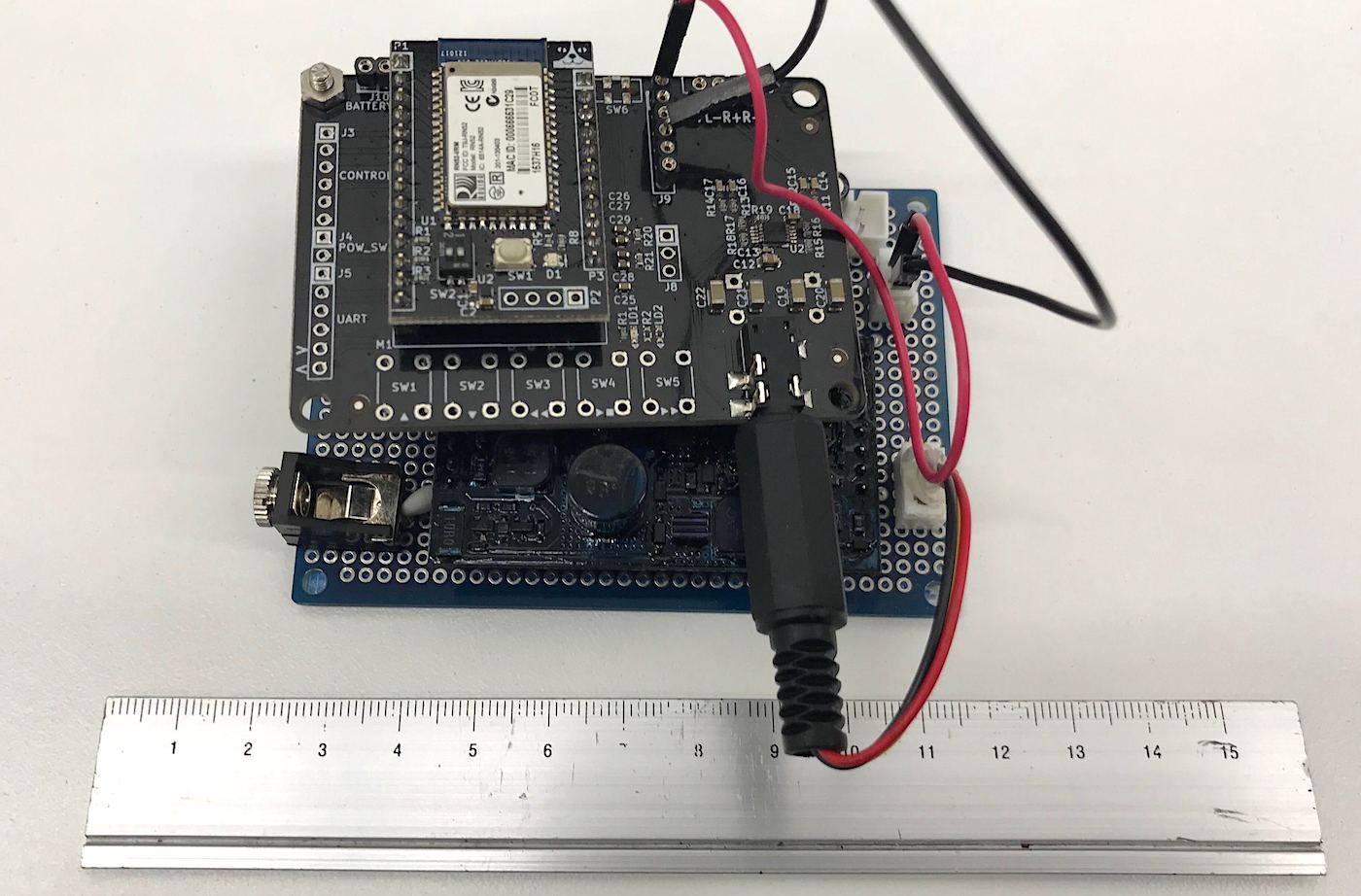}
\caption{The prototype design of {\it wavEMS}. }
\label{fig:wavems}
\end{figure}

We named our system ``{\it wavEMS},'' and the system is shown in Figure~\ref{fig:system} and~\ref{fig:wavems}. 

The implementation consists of a Bluetooth module, piezo amplifier, battery, and voltage converters. RN-52 (Microchip Technology) allows the toolkit to connect to mobile devices or computers, and receives the audio signals. A 3.7 V Li-ion battery is used, which is converted to 12 V and $\pm$ 5 V with DC-DC converters. The audio signals and the converted power source activates a PWM amplifier (IFJM-001, Marutsuelec Co., Ltd.), which outputs EMS signals to the electrodes (the maximum voltage is $\pm$ 60 V, and the current is limited to a maximum of 100 mA). For the software, we prepared a Processing program that can output typical EMS waveforms. However, for {\it wavEMS}, any type of audio can be used for stimulation output. {\it wavEMS} allows to output various waveforms including; sine waves, triangle waves, square waves, Russian current, and other desired waveforms. Waveforms of EMS can influence the experience in many ways. Another important benefit of {\it wavEMS}, is where it is controlled wirelessly. This helps in improving the mobility of EMS prototyping.

One negative aspect of {\it wavEMS} is where it can output signals that are not usually used for EMS, and can be {\it dangerous} without knowledge. Therefore, when using {\it wavEMS}, understandings of safety will be more strictly required. We recommend going through safety guidelines~\cite{tochiguide}, and to check the considerable threats, especially the effect of the frequency.

\begin{figure}[]
\centering
  \includegraphics[width=1.00\columnwidth]{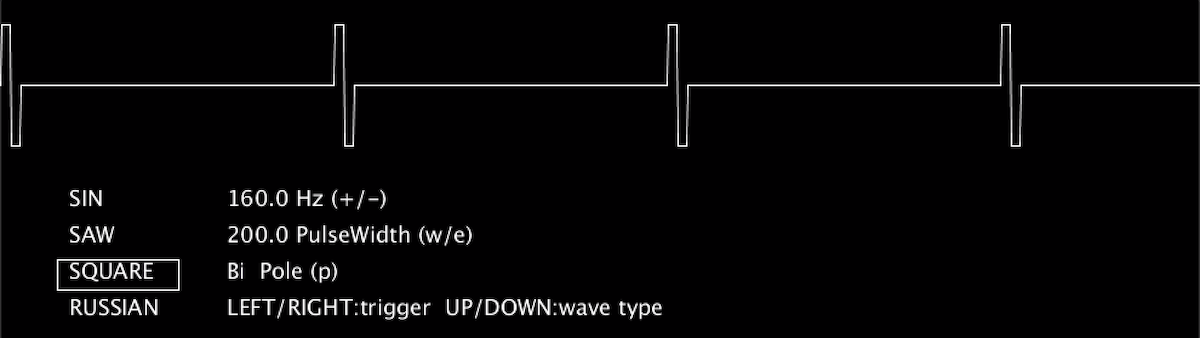}
  \caption{The user interface. Users can choose and manipulate the output waveforms.}
      \label{fig:ui}
\end{figure}

\subsection{How to Use}
The audio output from the computer can be used for {\it wavEMS}. For prototyping, we provide a user interface which is designed based on the Minim library\footnote{http://code.compartmental.net/minim/} of Processing\footnote{https://processing.org/}. In this user interface, the user can choose the output waveform from sine, saw, square waves of Russian current, as well as the frequency and pulse width (Figure~\ref{fig:ui}). This allows users to try popular types of outputs, which are considered safe and effective. The controller of {\it wavEMS} is not limited to Processing, and other tools such as Max/MSP can be used. Furthermore, audios from other resources can be used (e.g., Youtube), and can be used to feel the sound output of various audio. However, we must note that using audio signals can sometimes become dangerous. Therefore, it is strongly recommended to follow safety guides~\cite{tochiguide}, and not to attach electrodes to uncommon body parts (around or close to the chest).


\subsection{Performance}
The prototype of {\it wavEMS} was tested by the author. Frequencies were adjusted between 20--400 Hz and pulse widths were adjusted between 40--600 $\mu$s. The system successfully provided outputs with the desired parameters, with the desired waveform (both monophasic and biphasic), and appropriate muscular contraction was observed. However, due to characteristics of the filter of the system, the intensity tended to differ depending on the waveform, when the audio output was set to a fixed level on the computer used for the signal generation. Currently, the intensity tended to be stronger for square waves, and lower for triangular waves. Therefore, we set the gain of through the Minim library for square, sine, and triangular waves at -2, 2, 6 respectively via software. However, it is still recommended to adjust these gain parameters more precisely, when comparing the effect of the waveforms at the same intensity. 

\begin{figure}[h]
\centering
  \includegraphics[width=1.00\columnwidth]{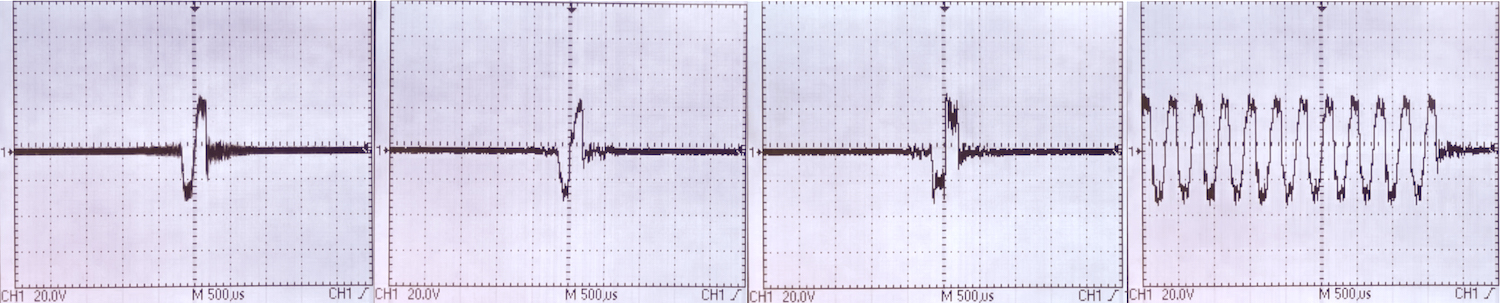}
  \caption{The waves are set to biphasic 160~Hz, 120 $\mu$s pulse width. They are sine, triangular, squared and Russian current output respectively.}
      \label{fig:oscillo}
\end{figure}

Figure~\ref{fig:oscillo} shows the output of {\it wavEMS} observed through an oscilloscope, with a 10 k$\Omega$ resistor. We are aware that there are noises on the waves, however, due to the nature of human perception to electricity, we believe that high-frequency noises are not perceptual and have only a minor effect to the EMS experience. 

\section{Discussion}
When experiencing stimulation from EMS systems, it is common to use signals around 1--150 Hz and pulse width with 30--800 $\mu$s~\cite{tochiguide}. Signals around this area can be usually effective for muscle actuation purposes. However, it is known that signals at high-frequencies are more difficult to be perceived by the human body. Therefore, using audio signals with high frequencies may not be effective for muscle stimulating purposes. In addition, risks for heat generation can become more noticeable in such situation, thus should not be used for a long duration continuously.

As a limitation of our current system, the system only allows a single channel output. Furthermore, there are still areas of improvements regarding energy efficiency. We plan to improve these elements and to provide the system as an opensource toolkit. 

The bus voltage can be controlled by the resistance of the resistor included in the circuit, and lower bus voltage can improve the reactive power and electromagnetic interference. For our initial testing, we use a 3.9 k$\Omega$ resistor for the amplifier to adjust the bus voltage (this allows to emit the maximum voltage at $\pm$ 60 V). However, this amplitude may be too strong for many users, and typically, lower voltage can be enough for HCI purposes. Therefore, this resistor can be replaced with a large resistance, which can contribute to energy efficiency as well.

For typical square pulsed based EMS systems, galvanic isolation between the power source side and the signal generation side are applied to improve safety. In our current implementation, the board is isolated with the signal generator (e.g., computer) by using Bluetooth communication, however, it still does not have isolation between the circuits of the signal receiver and the power source. Therefore, users must keep in mind not to touch the system while it is active, or otherwise should be covered with insulating materials.

\section{Conclusion}
This paper introduces {\it wavEMS}, an audio-based EMS system for prototyping and research explorations. While many HCI researchers use squared pulsed-waves, there are also potential to use other variations of waveforms, which may lead to different muscular effects and skin perception. Therefore, it is important for us to extend the possibility of the waveforms of EMS for further research development in HCI. In order to provide such opportunity, we developed {\it wavEMS}, which consists of a Bluetooth module and an amplifier for audio signals. A user interface was developed, so that researchers can test variations of typical EMS waveforms with various parameters. Our system allows improving the signal variation freedom of electrical stimulation. We believe that this paper can help improve HCI applications, and to open up new application areas.